\documentclass[11pt]{article}
\usepackage{adjustbox}
\usepackage{cite}
\usepackage{framed}
\usepackage{tikz,graphicx,graphics,color}
\usepackage{amsmath,amssymb,amsfonts}
\newcommand{\rem}[1]{}

\newsavebox{\astrutbox}
\sbox{\astrutbox}{\rule[-5pt]{0pt}{20pt}}

\newcommand{\bwt}{\begin{widetext}}
\newcommand{\ewt}{\end{widetext}}

\DeclareMathAlphabet{\mathbi}{OML}{cmm}{b}{it}

\newcommand{\bel}{\begin{equation}\label}
\newcommand{\ee}{\end{equation}}
\newcommand{\bq}{\begin{quote}}
\newcommand{\eq}{\end{quote}}
\newcommand{\beq}{\begin{eqnarray}\label} 
\newcommand{\eeq}{\end{eqnarray}} 
\newcommand{\bc}{\begin{center}} 
\newcommand{\ec}{\end{center}} 
\newcommand{\ben}{\begin{enumerate}}
\newcommand{\een}{\end{enumerate}}
\newcommand{\bit}{\begin{itemize}}
\newcommand{\eit}{\end{itemize}}

\newcommand{\bpsi}{\boldsymbol{\psi}}
\newcommand{\shalf}{{\ensuremath{\scriptstyle\frac{1}{2}}}}

\newcommand{\dx}{{\mbox{\rm d}}}

\makeatletter
\@addtoreset{equation}{section}

\makeatother

\begin{document}

\bc
\textbf{\color{blue}\large Martin David Kruskal}
\par\smallskip
\textbf{\color{blue}September 28, 1925 -- December 26, 2006}
\par\smallskip
\textbf{\color{blue}Elected ForMemRS 1997}
\ec
\bc
\textbf{J. D. Gibbon}\footnote{j.d.gibbon@ic.ac.uk}
\par
\textbf{Department of Mathematics, Imperial College London, UK}
\par\bigskip
\textbf{S. C. Cowley}\footnote{steven.cowley@ccc.ox.ac.uk} \textbf{FRS}
\par
\textbf{Corpus Christi College, Oxford}
\par\bigskip
\textbf{N. Joshi}\footnote{nalini.joshi@sydney.edu.au}
\par
\textbf{School of Mathematics and Statistics, University of Sydney, Australia}
\par\smallskip
and 
\par\smallskip
\textbf{M. A. H. MacCallum}\footnote{m.a.h.maccallum@qml.ac.uk}
\par
\textbf{Department of Mathematics, Queen Mary University of London, UK}
\ec
\bc
\textbf{\color{red}To appear in Biographical Memoirs of Fellows of the Royal Society}
\ec
\par\vspace{0mm}\noindent
\begin{abstract}
Martin David Kruskal was one of the most versatile theoretical physicists of his generation 
and is distinguished for his enduring work in several different areas, most notably plasma 
physics, a memorable detour into relativity, and his pioneering work in nonlinear waves. In  
the latter, together with Norman Zabusky, he invented the concept of the soliton and, with 
others, developed its application to classes of partial differential equations of physical 
significance. 
\end{abstract}


\section{\large Introduction}

Imagine you are at a conference at a European venue enjoying a leisurely, silent breakfast with a few other participants. 
Suddenly the door opens and an older, senior man joins the group. He picks up a jar of jam and starts to interrogate it 
energetically. ``Why is it round and not square?'' he asks suddenly, addressing the jar and its constituent parts in the 
third person. ``Why is it not hexagonal or even octagonal? Why is the lid not thicker? Why is the label pink and not 
green?'' The questions come thick and fast\,: Why, why, \ldots? The jar, we might add, remained unruffled and refused 
to answer. This interrogation upset the equilibrium of the other occupants of the table to such a degree that one lady 
retorted ``Oh, I don't \textit{know}, Martin'' and left, at which point the occupants of the table returned to their 
original demeanour. Later that morning, with a shy smile of silent apology, Martin presented the lady with a beautifully constructed origami flower in an origami box which has been kept and treasured to this day (see Fig. 1). This anecdote 
is an illustration of Martin Kruskal's relentless desire to interrogate any problem or issue that caught his interest, 
turning it inside out and upside down, until he had thoroughly understood it, yet with a disarming charm that gained 
the affection of those who knew and worked with him. He was recognized by all as a kind and generous man who had a 
penetrating, hyperactive and insatiably curious intellect. Fundamentally, he just wanted to understand whatever was 
in front of him. 
\begin{figure}
\bc
\includegraphics[width=0.3\textwidth, angle =0 ]{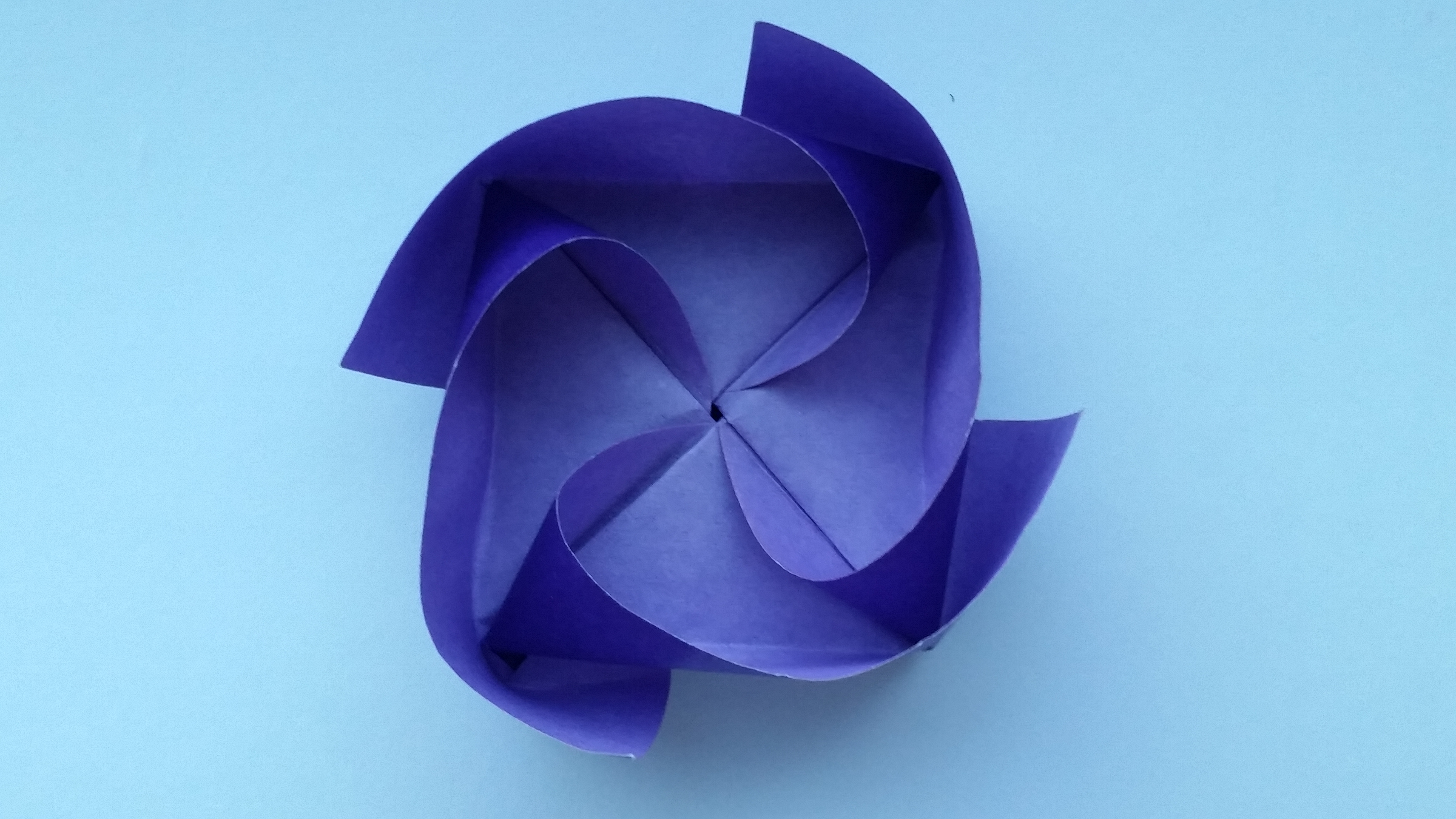}\qquad
\includegraphics[width=0.3\textwidth, angle =0 ]{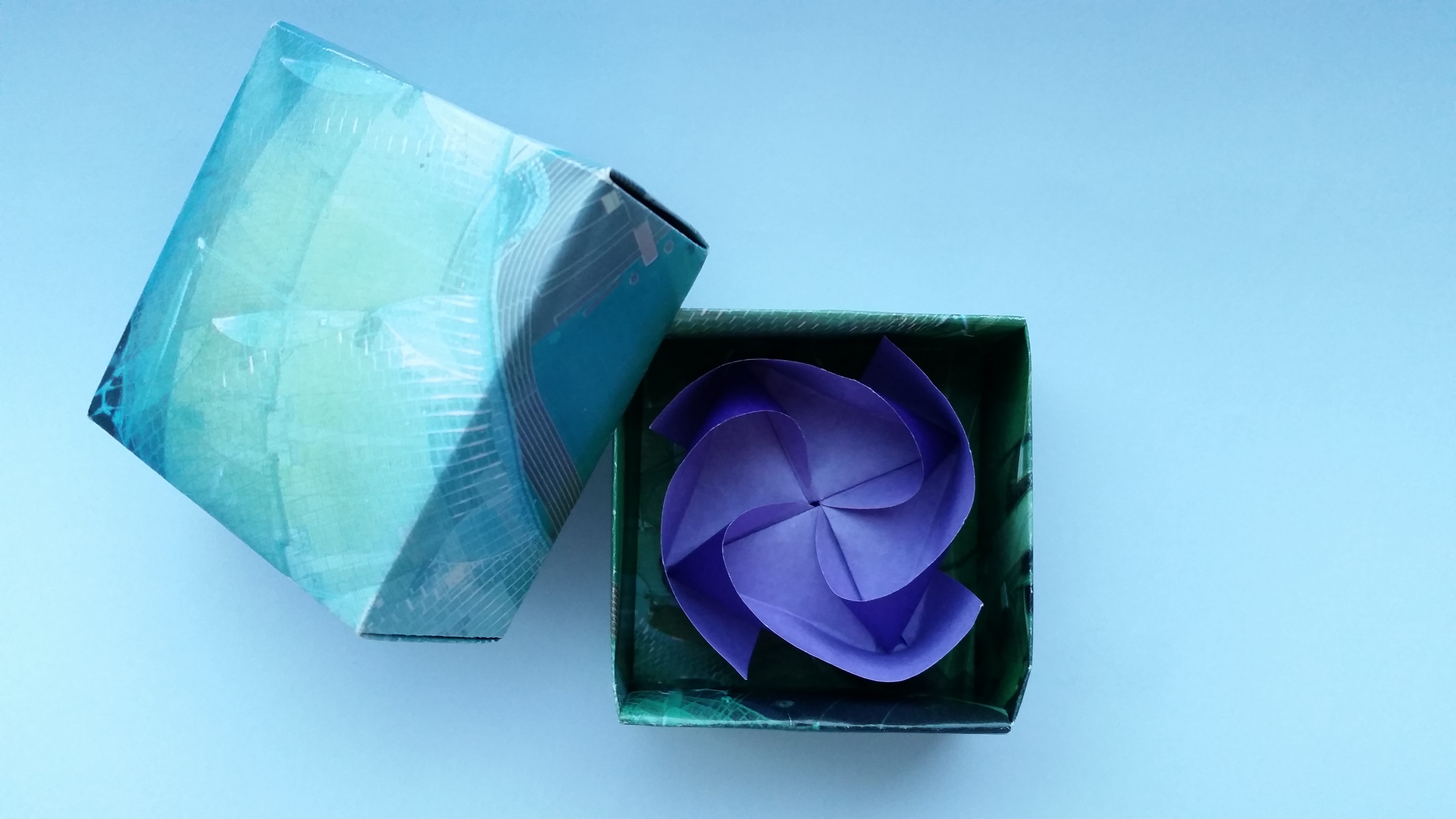}
\ec
\caption{\scriptsize An origami flower and its box, both made by Martin Kruskal and presented to Sheila Gibbon about 30 years ago.}
\end{figure}
\par\indent\noindent
A string of obituaries appeared on Martin David Kruskal's death in 2006 which contain a variety of facts about his early career\,: see the Los Angeles Times (2006), O'Connor and Robertson (2006), Eilbeck (2007), the New York Times (2007), SIAM (2007) and a more recently written memoir for the US National Academy of Sciences by Deift (2016). Martin was born in New York City on September 28, 1925 and grew up in New Rochelle, Westchester County, New York. He attended Fieldston High School in Riverdale, New York, and entered the University of Chicago from which he obtained his BS in 1945. Richard Courant persuaded Kruskal that he should undertake research at his new Institute, now the famed Courant Institute of Mathematical Sciences of New York University. Kruskal became an Assistant Instructor there in 1946 and, after studying for his MS, was awarded the degree in 1948. He then undertook research advised by Richard Courant and Bernard Friedman, during which time he married Laura (Lashinsky) in 1950. They had three children\,: Karen, Kerry, and Clyde. He submitted his thesis ``The Bridge Theorem for Minimal Surfaces'' and was awarded his doctorate in 1952. In 1951 he moved to Princeton where he had taken up a post in Project Matterhorn, which was re-named Princeton Plasma Physics Laboratory after declassification in 1961. In 1961 he became a Professor of Astronomy, then a founder and chair of the Program in Applied and Computational Mathematics (1968) and then, following that, he became a Professor of Mathematics (1979). He retired from Princeton in 1989 and joined the Mathematics Department of Rutgers University, holding the David Hilbert Chair of Mathematics. Professionally he was always known as Martin, but was always called David by his wife Laura and his family. His father, Joseph B. Kruskal, Sr., was a fur wholesaler and his mother, Lillian Kruskal Oppenheimer, founded the Origami Center of America in New York City (Origami USA). He was one of five children. His two brothers were Joseph Kruskal (1928-2010), discoverer of multi-dimensional scaling, the Kruskal tree theorem and Kruskal's algorithm in computer science, and William Kruskal (1919-2005), a statistician known for the Kruskal-Wallis test.
 
\par\smallskip
Following his mother, Martin had a great love of origami, games, puzzles and symmetries. The web has many entries regarding his card trick called the Kruskal count (AMS 2016). Laura once said that they originally met at his mother's Origami Center in NYC and Laura herself was equally well known as a lecturer and writer on that subject. In later years they travelled widely together to many scientific meetings at which Laura would organize anyone in reach into an origami group, whether they wished to or not! They were both much loved characters, larger than life, who made things happen and made any gathering or dinner a fun place to be. One story is that on a visit to a meeting in Rome, Laura sat on a bench in a public garden. As was her habit, she struck up a conversation with another inhabitant of the bench. She soon discovered that he was a Vatican official who was persuaded to get her an audience with the Pope. We have no corroboration of this story, but if it isn't true it ought to be, because it entirely reflected her character. Whether she managed to teach origami to the Pope is unknown but she no doubt tried.
\begin{figure}%
\bc
\ec
\caption{\scriptsize \textbf{\color{blue}The cartoon drawn by Leo Cullum and printed in the version of this article in 
Biographical Memoirs of Members of the Royal Society has been reproduced by permission of the Cullum family.}}
\end{figure}
\par\smallskip
The scale of the modern sciences is so huge that results that have taken so much of our own time and labour seem a mere grain of sand on a vast beach. Very few of us achieve results that leave marks on the coastline or even a few short-lived ripples on the sand. Martin was a rare example of one whose talents across a broad range of subjects have left the scientific coastline permanently changed. He was an old-school scientist who pursued a career `lifestyle' that deliberately rejected the devotion of so much time to the writing of grant proposals and the pursuit of influence through service on committees. His career ran against the modern trend in which research has increasingly become an industry, where goals, pathways to objectives and ultimate outcomes dominate a scientist's life and the acquisition of grant money its main concern. Instead, he spent his time in the pursuit of ideas and in the encouragement of others, particularly young scientists. Although it is reputed that he never had a grant, he was so successful that the absence of grant money was no barrier to travel because he was in such demand\,: conferences tended to come alive when he was in attendance. From the 1970's onwards he travelled the world extensively, often accompanied by Laura. In his capacious bag he carried a vast array of `just in case I might need them' articles, including a large range of materials and implements for his origami. Darryl Holm, now of Imperial College London, tells the story that in the Australian house of one of the co-authors of this memoir (Nalini Joshi), he used a word the meaning of which Martin felt was contextually wrong. After some discussion Nalini supplied them with two dictionaries, one of which supported Darryl and the other Martin. ``OK'' said Martin, ``In my bag I always carry 3 dictionaries so let's check in those.'' In the end Martin won 3-2. As much as anything the story illustrates his enduring passion for precise definitions. He also insisted on logical thought. Whenever a graduate student was stuck on a technical point, his exhortation to them was always ``Follow the logic!''. 
\par\smallskip
He was elected a Member of the US National Academy of Sciences in 1980, the American Academy of Arts and Sciences in 1983, a Foreign Member of the Royal Society (ForMemRS) in 1997, a Foreign member of the Russian Academy of Arts and Sciences in 2000 
and an Honorary Fellow of the Royal Society of Edinburgh (FRSE) in 2001. In 2000 he was also awarded an honorary doctorate by Heriot-Watt University, Edinburgh. Among his many prizes, he was awarded the National Medal of Science in 1993 and, together with Gardner, Greene and Miura, the Steele Prize by the American Mathematical Society in 2006. Listed alphabetically among his PhD students are Ovidiu and Rodica Costin, Jishan He, Nalini Joshi, Robert Mackay FRS (jointly with John Greene), Steven Orszag and G. V. Ramanathan. His work spanned many fields including major contributions to plasma physics, relativity, what are now called `integrable systems', and a lifelong interest in both asymptotology (his label) and surreal numbers. 


\section{\large Contributions to Plasma Physics and Fusion}\label{sect:plasma}

In 1951 Martin was recruited by Lyman Spitzer\footnote{Lyman Spitzer (1914-1997) was a reknowned astro-physicist who made major contributions to interstellar and plasma physics, space astronomy and nuclear fusion.  After WW2 he became Director of Princeton's Observatory and the Director of the classified Matterhorn project. He was the first to suggest the placing of a telescope in space and was a major force behind the development of the Hubble Space Telescope. In 2003 NASA launched an infra-red space observatory and named it the Spitzer Space Telescope in his honour.} to join the then classified controlled fusion project at Princeton called Project Matterhorn\footnote{This section was written by Steven Cowley FRS.}. This project pursued the idea of magnetic confinement of a fusion plasma by three dimensional magnetic fields, a concept named the Stellarator by Spitzer.  Martin was Spitzer's first employee and an inspired choice, as it turned out.  Almost nothing was known about plasma physics in 1951 and certainly magnetic confinement theory was nonexistent.  It was, therefore, a perfect time to enter the field and Martin took full advantage of the opportunity to help define and develop a new area of physics.  Spitzer asked Martin to look at a mathematical problem while waiting for his security clearance to come through.   Specifically he asked whether they could make integrable magnetic fields that lie on surfaces in three dimensions?  Spitzer wanted to make sure that the field lines in the Stellarator would stay confined since the hot plasma ions and electrons would follow the field lines -- it is not clear how much of this motivation Spitzer could tell Martin until he was cleared.  In an unpublished report (Kruskal 1952) Martin proved that for small rotational transform, $\iota \ll 1$, the deviation from perfect magnetic surfaces was beyond all orders in powers of $\iota$.  Clearly Martin was already thinking deeply about \textit{integrability} and \textit{asymptotics beyond-all-orders} in 1951 (although admittedly in a simplified system), well before, for example, Kolmogorov's famous 1954 paper on the existence of surfaces in Hamiltonian systems (Kolmogorov 1954). Spitzer's question remains central to Stellarator research. Indeed, modern Stellarators are designed by iteratively searching for integrable fields (see e.g. Helander 2014).

A necessary condition for a practical fusion system is hydromagnetic stability.  Martin made many major contributions to the understanding and formulation of plasma stability.  In 1954 Kruskal and Martin Schwarzschild published one of the earliest calculations of plasma stability (Kruskal and Schwarzschild 1954). They treated the instability of a plasma supported against gravity by a magnetic field now commonly called the \textit{Kruskal-Schwarzschild Instability} and the \textit{kink} stability of a current carrying plasma column.  The largest stable plasma current in cylinder is set by the \textit{Kruskal-Shafranov Limit}. Martin calculated this limit in an initially classified report (Kruskal 1954) and the result was finally published in Kruskal, Johnson, Gottlieb and Goldman (1958). 

By the mid 1950s Spitzer had assembled a team of highly talented young theoreticians that produced a string of classic papers that have huge influence to this day.  Martin was very much at the centre of this effort -- we can only highlight the most important of his contributions.  With Russell Kulsrud he formulated the problem of the equilibrium of a magnetically confined plasma and showed that it can be obtained as a stationary variation of the energy (Kruskal and Kulsrud 1958).  In the famous \textit{Energy Principle} paper, Martin and three others of the Princeton group, Ira Bernstein, Ed Frieman and Russell Kulsrud, showed that positivity of the second variation of the magnetohydrodynamic (MHD) energy is a necessary and sufficient condition for MHD stability (Bernstein, Frieman, Kruskal and Kulsrud 1958).  They also showed how the principle could be used to calculate stability in complicated geometries.  The \textit{Energy Principle} is the basis of most modern MHD stability calculations.  Particle collisions in fusion plasmas are rare and therefore the MHD fluid equations are an inaccurate description of the plasma behaviour.   Martin and Carl Oberman developed the first kinetic energy principle (Kruskal and Oberman 1958) based on the collisionless guiding centre description of the plasma -- this paper too is the starting point of many modern calculations.
    
The adiabatic invariants of the guiding centre approximation for charged particles in a magnetic field are represented by an asymptotic series in the small parameter $\epsilon$ where $\epsilon$ equals the Larmor radius divided by the scale length of the magnetic field.  Martin became fascinated by the structure and generality of this problem, in particular showing that invariance could be proved to all orders in $\epsilon$.  Russell Kulsrud proved first that the adiabatic invariant of the Harmonic Oscillator is invariant to all orders in ${\dot\omega}/\omega^{2}\ll 0$ where $\omega(t)$ is the instantaneous oscillation frequency, with the dot representing a time derivative (Kulsrud 1957). Martin demonstrated the same result for the first adiabatic invariant of a particle in a magnetic field, then in greater generality for an autonomous system of differential equations with all solutions nearly periodic (Kruskal 1962a).  When Martin presented the results of this paper to the Princeton group the seminar lasted for two days -- most of the audience stayed for it all!   

Plasmas support nonlinear electrostatic waves that do not undergo Landau damping. These were discovered by Ira Bernstein, John Greene and Martin 
(Bernstein, Greene and Kruskal 1957) and are usually referred to as BGK modes.  They showed that self-consistent solutions can be found in one dimension by adding the appropriate distribution of particles trapped in the electrostatic potential.  The dynamics of nonlinear collisionless waves in plasmas remains a vibrant area of research and BGK waves play a central role in our current understanding.

Martin's interest in plasma physics and fusion research waned in the 1960s as he became more consumed by the study of nonlinear PDEs and in particular the KdV equation\,: see \S\ref{sect:KdV}.  Nonetheless, the influence of Martin's mathematical style is imprinted on modern plasma physics -- perhaps most obviously on those of us brought up on his notions of \textit{Asymptotology} (Kruskal 1963) -- but what is perhaps more surprising is that Martin could, and often did, think like a physicist, especially when he worked with experimentalists.  He was not afraid to use intuition when rigour was unavailable and he delighted in cartoon explanations of physical processes.  It is rare indeed that someone looks like a physicist to physicists and a mathematician to mathematicians!


\section{\large Kruskal's contribution to Relativity}\label{sect:rel}

In 1960, papers by Kruskal (1960), and independently by Szekeres (1960), found the maximal analytic extension 
of Karl Schwarzschild's\footnote{Karl Schwarzschild (1873--1916) was the father of Martin Schwarzschild (1912--1997) 
who was himself a co-author with Martin Kruskal -- see \S\ref{sect:plasma}.} vacuum solution, and coordinates for 
it\footnote{This section was written by Malcolm MacCallum.}. The metric is now well understood to represent a 
spherically symmetric black hole. Its structure, horizon and singularities are explained in introductory courses 
and texts by making use of the Kruskal-Szekeres form, which is known thus to all students of the theory. 
\par\smallskip
Kruskal's paper was rather unusual in that it was actually written by Wheeler (Wheeler and Ford 
1998, pages 295-296). Kruskal had shown Wheeler his results (allegedly on a napkin in a lunchroom) some time in 
1956-7: Charles Misner recalls being told about them in 1958 and Kruskal's paper says Wheeler described them at 
a 1959 conference. Prompted (as described in Wheeler and Ford 1998) by work of Misner and others, Wheeler wrote 
the results up and submitted the paper without telling Kruskal (though over his name)\,; the first Kruskal knew 
of it was when he received the galley proofs. Wheeler records that Kruskal ``was mystified only briefly'' and 
suggested it be published as joint work, but Wheeler demurred on the grounds that all the important ideas belonged 
to Kruskal. 
\par\smallskip
The starting point was the metric given by Karl Schwarzschild (Schwarzschild 1916), only seven weeks after Einstein's 
paper presenting the final form of his theory. In what are now called Schwarzschild coordinates, it reads
\begin{equation} \label{Schw}
\dx s^2 = - (1-2m/r)\dx t^2 + \dx r^2/(1-2m/r) + r^2 (\dx \theta^2+\sin^2 \theta\dx \phi^2)\,,
\end{equation}
where $m$ is the gravitational mass that would be measured by a far away observer, in geometrized units, and $4\pi r^2$ 
gives the areas of the spheres of symmetry. 
\par\smallskip
The form \eqref{Schw} is valid in $r>2m$, or in $r<2m$, but not at $r=2m$. That surface is now understood to be 
the black hole horizon, a light-like surface bounding the black hole region from which light cannot escape\,: the 
coordinates of \eqref{Schw} are singular there. An imperfect understanding of such coordinate singularities (still 
sometimes found) be-devilled early interpretations (Eisenstaedt 1982), but Lema\^{i}tre (1933) wrote that ``the 
singularity of the field is not real and arises simply because one wanted to use coordinates for which the field 
is static''. Others had found the same. The Eddington-Finkelstein coordinates, which go smoothly across the horizon 
(Eddington 1924, Finkelstein 1958), clarified the matter by using a null (light-like) coordinate like those used by 
Kruskal.  This work still did not provide the maximal extension. 
\par
\begin{figure}[h]
\begin{center}
\includegraphics[scale=0.6]{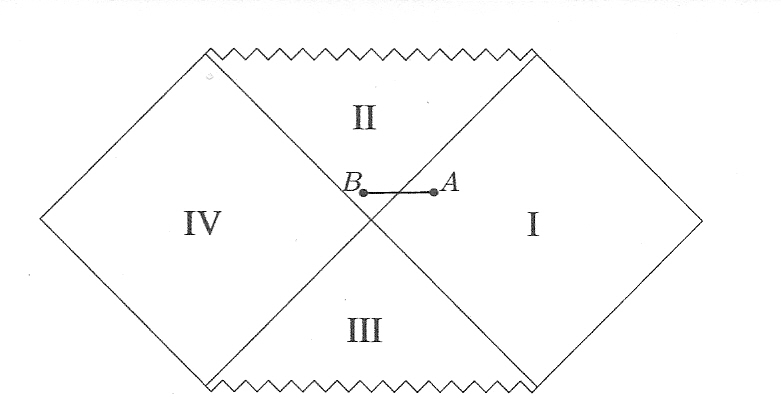}
\end{center}
\caption{\scriptsize Conformal diagram of the Kruskal-Szekeres maximal extension of Schwarzschild's solution. 
}\label{Penrose}
\end{figure}

\begin{figure}[h]
\begin{center}
\includegraphics[scale=0.3]{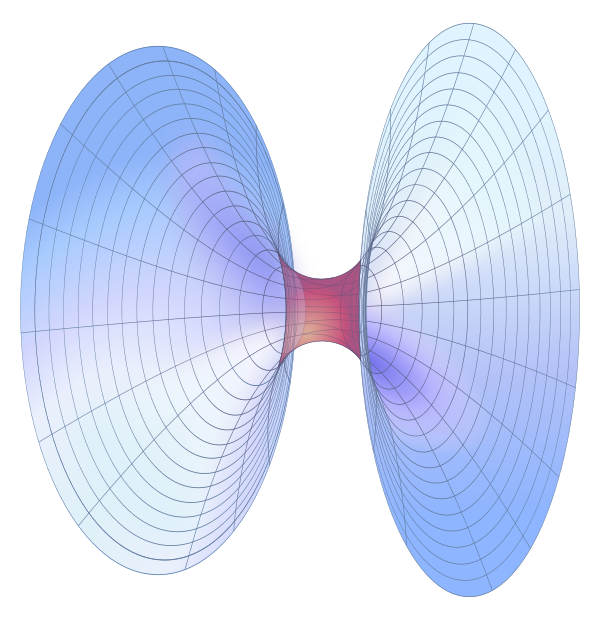}
\end{center}
\caption{\scriptsize  Diagram of a Schwarzschild solution wormhole (from Wikipedia, contributor Kes47)}\label{worm}
\end{figure}

\smallskip
That extension can be illustrated by its conformal \textit{Penrose diagram}\,; see Fig \ref{Penrose}. Such diagrams 
depict the metric transformed by a conformal factor $\Omega$, i.e.\ they show $\Omega^2$ times the original metric 
where the function $\Omega\rightarrow 0$ at infinity in such a way that the original space is mapped to a finite 
region. In this figure each point represents a sphere,  the coordinates $\theta$ and $\phi$ being suppressed, and 
light travels on lines at $45^o$ to the axes. Time directions are \textit{vertical} (i.e.\ at more than $45^o$) and 
space directions horizontal in the picture. The region labelled I is the region $r>2m$ in \eqref{Schw}. Region II 
($r<2m$) is the interior of the black hole, the horizon being the boundary between I and II. The jagged lines 
represent the future and past (true, not coordinate) singularities at $r=0$ and the left and right hand edges of 
the figure consist of points at infinity.
The region III is a white hole, from which light can emerge but not travel into. Intriguingly, region IV is a second exterior region. The line AB in the figure represents a three-dimensional surface composed of spheres whose size decreases to a minimum as one moves from A towards B. If one continues the line into region IV the spheres' sizes increase again. Representing each sphere by a circle, the wormhole can be drawn as in Fig. \ref{worm}. One might then imagine, as described in Kruskal's (or Wheeler's) paper, that such a wormhole could join two areas in the same space-time, although this is impossible in the Schwarzschild maximal extension itself.
Sadly for science fiction writers, such wormholes in the Schwarzschild space-time, being space-like, can only be 
traversed if travelling faster than light. However, there are other wormhole solutions which, although perhaps 
not present in nature due to the need for exotic forms of matter to sustain them, give very interesting possibilities,  
including causality violations. An excellent semi-popular account of these, which describes results from later 
technical papers, appears in chapter 14 of Thorne (1994).
\par\smallskip
One can arrive at Fig. \ref{Penrose} as follows. Kruskal and Szekeres used the coordinates
\begin{eqnarray}\label{krusszek}
  u &=& (r/2m - 1)^{1/2}e^{r/4m}\cosh (t/4m)\,,\\
  v &=& (r/2m - 1)^{1/2}e^{r/4m}\sinh (t/4m)\,,
  \end{eqnarray}
 in terms of which the metric becomes
  \begin{equation}\label{schwks}
   \dx s^2  =  32m^3 (\dx u^2  - \dx v^2 )/re^{r/2m}+ r^2 (\dx \theta^2  + \sin^2 \theta\, \dx \phi^2 )\,,
  \end{equation}
 with $r$ given in terms of the new coordinates implicitly by
 \begin{equation}\label{kstrans2}
   u^2  - v^2  = e^{r/2m} (r/2m - 1)\,.
 \end{equation}
Kruskal's paper includes a nice diagram showing the four regions. The coordinates of \eqref{schwks} and coordinates 
covering Fig. \ref{Penrose} showing the conformally transformed metric, are related by
 \begin{equation}
 u' = \arctan(u/\sqrt{2m}), \qquad v' = \arctan(v/\sqrt{2m})\,.
 \end{equation}
In Fig. \ref{Penrose}, $-\shalf \pi < u' < \shalf \pi$, $-\shalf \pi < v' < \shalf \pi$ and $-\shalf \pi < u' +v' < \shalf \pi$\,.
\par\smallskip
 As well as prompting the intriguing work on wormholes, the results of Kruskal and Szekeres stimulated comprehensive 
investigations of the structures of other black holes (see Carter 1973) and helped promote the uses of global analysis 
that has led to work on singularity theorems and other issues (Hawking and Ellis 1973) that still continues.

\section{\large The KdV equation and integrable nonlinear systems}\label{sect:KdV}

It is hard for younger scientists to imagine the scientific world of the 1950's with prehistoric computational facilities 
where \textit{linear} thinking still dominated\footnote{This section was written by John Gibbon.}. The large-scale PDE 
systems of applied mathematics and theoretical physics appeared so intractable that the reflex response was to ask what 
the linear approximation gave. Physics and applied mathematics have always abounded with special solutions of systems 
that are intrinsically nonlinear but the effect that even small nonlinearities could have on a predominantly linear 
system was not at all well understood, nor did the question necessarily spring to mind.  An argument has often been 
made that the emergence of quantum mechanics in the 1920s, an essentially linear science in that decade, side-lined 
the knowledge accumulated on 19th century nonlinear differential equations, which had dropped out of fashion. There 
may be something in this view but the massive disruptions of the two world wars would 
suggest that there were many more factors involved than just this. There also existed people whose thinking was ahead of 
their time. For instance, T. H. R. Skyrme came up with the idea of the sine-Gordon equation\footnote{Subscripts denote 
partial derivatives.} (Skyrme 1958) 
\bel{sge1}
\phi_{xx} - \phi_{tt} = m^{2}\sin\phi
\ee
as a model nonlinear field theory for strong interactions, a step beyond the linear Klein-Gordon equation. The idea was 
to write down a fully \textit{nonlinear field equation} which has local solutions of finite energy which cannot be reached 
by perturbation theory. \eqref{sge1} has a travelling-wave solution, called a kink, of the form
\bel{sge2}
\phi (x,\,t) = 4\arctan \exp \left\{{m\gamma (x-vt)+\delta}\right\} \qquad\qquad \gamma ^{2}= (1-v^{2})^{-1}
\ee
where $\delta$ is an arbitrary phase shift and $v$ the kink velocity\footnote{Kinks (anti-kinks) are solutions that move 
$\phi$ from $0$ to $2\pi$ (and vice-versa). It is the profiles of the  partial derivatives $\phi_{x}$ or $\phi_{t}$ that 
are the solitons. Thus Perring and Skyrme (1962) had found the first analytical multiple soliton solution but there was 
no hint in the paper of deeper properties that underlay this result.}.  Perring and Skyrme (1962) actually constructed 
an exact double kink solution of \eqref{sge1} with equal but opposite velocities fired at one another as particles from 
$-\infty$ and $\infty$. These merged and then emerged intact. Neither the importance of the model nor the significance 
of their solution was recognized at the time. 
\par\smallskip
A set of methods began to be developed in the early 1960's, which sought to determine how dispersion or dissipation 
balanced the nonlinear terms in PDEs on some set of stretched time and space scales determined by a small amplitude 
parameter $\varepsilon$. Different names have been used depending on the circumstances but the methods of reductive 
perturbation theory, stretched co-ordinates and multiple scales are names that will be familiar to those who have 
worked on weakly nonlinear systems. For instance, Stuart (1960) was the first to develop these ideas for plane 
Poiseuille flow at the point of critical instability of the linear system on a time scale $T=\varepsilon^{2}t$. With 
the inclusion of a space variable $X=\varepsilon x$ these ideas were developed further in a series of papers by Benney 
and Newell (1967), Newell and Whitehead (1969) and Newell (1974) for fluid convection, and Stewartson and Stuart (1971) 
for plane Poiseuille flow. For predominantly dispersive systems, such as those found in plasmas, nonlinear chains and 
surface water waves, reductive perturbation methods that use an amplitude $\varepsilon u(x,t)$ on time and space scales 
$\xi = \varepsilon^{q/2}(x-c_{p}t)$ and $\tau = \varepsilon^{3q/2}t$ give PDEs of the type (see Dodd, Eilbeck, Gibbon 
and Morris 1982)
\bel{red1}
u_{\xi\xi\xi} + 6u^{q}u_{\xi} + u_{\tau} = 0\,.
\ee
$q=1$ is the KdV equation\footnote{The coefficient of the $uu_{\xi}$-term can be altered at will by a re-scaling of $u$ 
so in this memoir we will freely use different values appropriate to the occasion.} while $q=2$ is the modified KdV (mKdV) equation. The KdV equation was already known, and received its name, from a paper by Korteweg and de Vries (1895) on the dynamics of small surface waves in shallow water. The solitary wave, observed and named by Scott Russell (1844) (but not 
the solution \eqref{red2}), shows up in the solution 
\bel{red2}
u = \shalf a^{2}\mbox{sech}^{2}(ax-a^{3}t + \delta)\,.
\ee
Among many examples, waves in cold ion plasmas obey the KdV equation -- see Washimi and Taniuiti (1966). It was this logical development of thought that ultimately led to the KdV or mKdV equations as PDEs that described weakly nonlinear behaviour of what appeared to be intractable bigger systems. 

\subsection{\small The work of Zabusky and Kruskal and the FPU problem}\label{ZKFPU}

A very early example of the investigation of the effects of nonlinearity was made at Los Alamos National Laboratory by Fermi, Pasta and Ulam (1955) who were interested in the behaviour of systems which were primarily linear but into which nonlinearity was introduced as a perturbation. In the absence of such perturbations, the energy in each of the normal modes of the linear system would be constant. It was expected that the nonlinear interactions between the modes would lead to the energy of the system being evenly distributed throughout all of the modes\,: a result which would be in accordance with the equipartition theorem. The results they obtained contradicted this idea. The importance of what is now called the FPU problem is that the unexpected nature of their results stimulated work on these types of nonlinear systems and some of the modern work on solitons stemmed directly from it. A brief resume of the their original report is this\,: consider a dynamical system of $N$ identical particles of unit mass (FPU had $N =64$) on a line with fixed end points with forces acting between nearest neighbours. If 
$Q_{n}(t)$ denotes the displacement from equilibrium of the $n$th particle then the equation of motion for this particle can 
be written as
\bel{FPU1}
\ddot{Q}_{n} = f(Q_{n+1} - Q_{n}) - f(Q_{n} - Q_{n-1})\,.
\ee
Two examples of the choice FPU made for $f$ were either $f = \gamma Q + \alpha Q^{2}$ or $f = \gamma Q + \beta Q^{3}$, where 
$\gamma$ denoted the linear chain constant and the constants $\alpha$ and $\beta$ were chosen such that the maximum displacement of $Q_{n}$ was small. Using these two nonlinearities, FPU integrated equation \eqref{FPU1} numerically\footnote{It would appear that these computations were actually performed by a young woman named Mary Tsingou (Dauxois 2008).} on one of the earliest valve computers called MANIAC 1. Using initial data in the form of a sine-wave, they found that the energy did not spread throughout all the normal modes but remained in the initial mode and a few nearby modes. Furthermore the energy density of 
those nearby modes had an almost periodic behaviour in time. Over a large number of oscillations, the energy in each normal 
mode was seen to be almost periodic in time, with no loss of energy to higher modes as time increased. The precise explanation of this periodicity, which they called `recurrence', stimulated a deeper study of equations such as \eqref{FPU1}. In the continuum limit, equation \eqref{FPU1} can be transformed into the KdV equation by using the method of stretched co-ordinates mentioned earlier. The choice of $f (Q) = \exp (-Q)$ makes \eqref{FPU1} into the Toda lattice (Toda 1967). 
\par\smallskip
The word `soliton', after John Scott Russell's `solitary wave' (Scott Russell 1844), first appeared in the paper by Zabusky 
and Kruskal (1965). Kruskal had been interested in the FPU problem for some time, particularly in the explanation for why recurrence occurred\,: see his earlier paper on recurrence with respect to his plasma work (Kruskal 1962a). Together with 
Norman Zabusky of Bell Laboratories, a very experienced computational physicist, Kruskal described a numerical study of the 
KdV equation (Zabusky and Kruskal 1965) in a form where we revert to $(x,\,t)$ co-ordinates
\bel{ZK1}
u_{t} + uu_{x} + \delta^{2}u_{xxx} = 0\,.
\ee
It is noteworthy that when $\delta = 0$ the simple PDE $u_{t} + uu_{x} = 0$ causes waves to steepen in regions of negative gradient, ultimately causing a shock. Adding the smoothing dissipative term $\delta^{2}u_{xx}$ to the right hand side gives Burgers' equation. The addition of $\delta^{2}u_{xxx}$ to the left hand side disperses waves. In (\ref{ZK1}) they chose 
$\delta = 0.022$ with boundary conditions that were periodic such that $u(x,t) = u(x+2,t)$, with an initial condition $u(x,0) 
= \cos x$. They noticed that initially the wave steepened when it had a negative slope, which was a consequence of the dominance of the nonlinearity over the very small dispersive term, but once the wave had steepened the $\delta^{2}u_{xxx}$-term became important and balanced the nonlinearity. On the left of the steepened region oscillations developed each of which grew and reached a steady but different amplitude with each becoming a solitary wave in shape like \eqref{red2}. \textit{The remarkable property of these was that they passed through one another with only a change of phase} as they went through the cycles of evolution forced by the periodic boundary conditions. This phase shift ensured that the initial state did not quite recur but nevertheless it came close to recurrence, as in the FPU problem. Zabusky and Kruskal called these solitary waves `solitons' because of their particle-like property. Subsequent work showed that the particle-like property of solitions is robust with a change of boundary conditions from periodic to the whole line. The particle-like behaviour intrigued Kruskal because the elastic collisional properties (with phase-shifting) reminded him of quantum mechanical scattering. N. J. Zabusky has wrote a retrospective account of this work (Zabusky 2005) in the year before Kruskal's death in 2006. 

\subsection{\small The seven papers of Gardner, Greene, Kruskal, Miura and other collaborators}

In the period 1965 to 1974 a series of seven papers were written by Kruskal and his associates. Kruskal was a co-author on four of them and not all four named authors appear on every paper. The first paper by Gardner, Greene, Kruskal and Miura (1967) is the key paper but no numerical labelling appears in the title\,: the next six were entitled ``Korteweg de Vries equation and generalizations I--VI\,: ....''. Number I is authored by Miura alone (Miura 1968), number III was authored by Su and Gardner (1971), which concerned the derivation of \eqref{kdv1a} and \eqref{kdv1b} and number IV was authored by Gardner alone (Gardner 1971). Number V has Zabusky as a co-author (Kruskal, Miura, Gardner and Zabusky 1970). All of them explored the properties of the KdV equation. 
\par\smallskip
Let the Korteweg de Vries (KdV) and modified Korteweg de Vries (mKdV) equations be written in the form
\bel{kdv1a}
u_{t} - 6 uu_{x} + u_{xxx} = 0\,,
\ee
\bel{kdv1b}
v_{t} - 6 v^{2}v_{x} + v_{xxx} = 0\,.
\ee
The coefficients of -6 are adjustable by scaling. Robert Miura (Miura 1968) discovered that if 
\bel{kdv2}
u = v_{x} + v^{2} 
\ee
then 
\bel{kdv3}
u_{t} - 6 uu_{x} + u_{xxx} = \left(\frac{\partial~}{\partial x} + 2v\right)(v_{t} - 6 v^{2}v_{x} + v_{xxx})\,.
\ee
Clearly, if $v$ satisfies \eqref{kdv1b} then $u$ satisfies \eqref{kdv1a}, but not necessarily vice-versa. In Gardner, 
Greene, Kruskal and Miura (1967) it was shown that the Riccati equation \eqref{kdv2}, now known as the Miura transformation, 
is exactly linearizable\,: choose $v= \psi_{x}/\psi$, and also note that the KdV equation is invariant under a Galilean transformation $x = x' + 6\lambda t'$; $t = t'$ and $u = u' -\lambda$. These turn \eqref{kdv2} into (dropping the primes) 
\bel{kdv4}
\left\{- \frac{\partial^{2}~}{\partial x^2} + u(x,t)\right\}\psi = \lambda \psi\,.
\ee
How does $\psi$ evolve in time? This is found by using $v = \psi_{x}/\psi$ in \eqref{kdv1b} and then \eqref{kdv4}
\bel{kdv5}
\left\{\frac{\partial~}{\partial t} + \frac{\partial^{3}~~}{\partial x^{3}} - 3 (u - \lambda)\frac{\partial~}{\partial x} \right\} \psi = f \psi\,,
\ee
with $f$ taken as a constant. 
Readers will immediately recognize \eqref{kdv4} as the Schr\"odinger equation of quantum mechanics with the KdV variable 
$u(x,t)$ as the potential and $\lambda$ as a \textit{constant energy eigenvalue}. In elementary quantum mechanics one is normally given a potential and asked to solve for $\psi$ together with the corresponding energy spectrum -- in fact, for 
initial data $u(x,0)$ this is exactly what one does. However, for $t > 0$ one has an \textit{inverse problem}\,: given 
asymptotic properties of $\psi$ at $x = \pm \infty$, with a constant spectrum $\lambda$ and scattering data, namely 
reflection and transmission coefficients $a(k,0)$ and $b(k,0)$, can we reconstruct the potential $u(x,t)$ for all $t > 0$? 
In the entirely different subject of scattering theory, the solution to this inverse problem had already been answered by Gel'fand and Levitan (1955) and Marchenko (1955). Gardner, Greene, Kruskal and Miura (1967) showed how to use this scattering machinery to reconstruct the potential $u(x,t)$ for all $t > 0$. 

\subsection{\small The Lax pair formulation}\label{sect:Lax}

In a classic paper, Peter Lax (1968) re-formulated the idea in a wider context and showed how to write down scattering 
problems for a much wider set of PDEs. Consider a spectral problem (in one dimensional space $x$) with differential 
operator denoted by $L$ together with a time dependence denoted by the operator $P$ 
\bel{lax1}
L\psi = \lambda(t) \psi\,,\qquad\qquad \psi_{t} = P\psi\,.
\ee
What is the condition on $P$ and $L$ such that $\lambda$ is constant in time? 
Simple differentiation with respect to $t$ and re-substitution shows that 
\bel{lax3}
L_{t} = PL - LP = [P,\,L]\,.
\ee
We have considerable freedom to choose $L$ and $P$ but let us begin with the symmetric Schrodinger operator in \eqref{kdv4}
\bel{lax4}
L = - \frac{\partial^{2}~}{\partial x^2} + u(x,t)\,,
\ee
then, writing $P$ as a third order anti-symmetric operator taken from \eqref{kdv5}, we find the potential $u(x,t)$ 
evolves according to the KdV equation \eqref{kdv1a}, as it should. This highlights the fact that while $u(x,t)$ is 
deforming with time, the spectrum $\lambda$ remains constant. This is called an \textit{iso-spectral deformation.}  
It can now easily be seen that one can play a game\,: for a fixed $L$, such as \eqref{lax4}, one can choose a hierarchy 
of anti-symmetric operators $P$ which yield a corresponding hierarchy of PDEs. Likewise one can vary $L$ and make it 
a third order operator or endow it with a matrix formulation. 
\par\smallskip
In the paper by Miura, Gardner and Kruskal (1968) they also recorded a series of higher conservation laws\,: the first 
three are the standard mass, energy and momentum while the fourth was found by Whitham (1965). The Hamiltonian structure 
of the KdV equation appeared in the context of a hierarchy of PDEs (see Gardner (1971) and Gardner \textit{et al} (1974)). 
They showed that a recursion relation exists between the $Q_{k}$ in the Hamiltonian formulation $\partial_{t}u = - 
\partial Q_{k}/\partial x$ and $\partial_{t} P_{k} = \delta H_{k}/\delta u$. Beginning with $Q_1 = u$ one can generate 
the infinite sequence $Q_{k}$. The KdV equation is the second in the hierarchy with an infinite sequence of conserved quantities. A full Hamiltonian analysis can be found in Zakharov and Faddeev (1971). 

\subsection{\small Zakharov and Shabat and the AKNS formulation}\label{subsect:AKNS}

A further boost to the subject came when the distinguished Russian plasma physicist V. E. Zakharov and co-workers began 
to work in this area. Zakharov and Shabat (1972) extended these ideas by finding a Lax pair for the so-called nonlinear 
Schr\"odinger (NLS) equation 
\bel{NLS1}
iq_{t} + q_{xx} \pm 2q|q|^{2} = 0 \,.
\ee
The effect of this work opened the possibilities that there could be many more nonlinear systems whose solutions have 
the same particle-like properties as the KdV and mKdV equations\footnote{In the meantime, Wadati (1972) had shown how 
to find the scattering problem for the mKdV equation.}. Ablowitz, Kaup, Newell and Segur (1973) came up with an elegant formulation of a more general scattering problem (see also Ablowitz and Segur 1981) 
\bel{AKNS1}
\frac{\partial\psi_{1}}{\partial x} + i\lambda\psi_{1} = q\psi_{2}\,,\qquad\qquad
\frac{\partial\psi_{2}}{\partial x} - i\lambda\psi_{2} = r\psi_{1}\,,
\ee
with a time dependence on the $(\psi_{1},\,\psi_{2})$
\bel{AKNS2}
\frac{\partial\psi_{1}}{\partial t} =  A\psi_{1} + B\psi_{2}\,,\qquad\qquad
\frac{\partial\psi_{2}}{\partial t} =  C\psi_{1} - A\psi_{2}\,.
\ee
\rem{This is like a factorization of the 2-channel Schr\"odinger system $\bpsi = (\psi_{1},\,\psi_{2})$
\bel{AKNS2}
\left( - \frac{\partial^{2}~}{\partial x^2} + \mathbf{V}\right)\bpsi = \lambda^{2}\bpsi 
\qquad\qquad 
\mathbf{V} = \left(
\begin{array}{cc}
qr & q_{x}\\
r_{x} & qr
\end{array}
\right)
\ee}
It is easy to compute the compatibility conditions between \eqref{AKNS1} and \eqref{AKNS2}, which are 
\bel{AKNS3}
A_{x} = qC - rB\qquad\qquad B_{x} +2i\lambda B = q_{t} - 2Aq \qquad\qquad C_{x} - 2i\lambda C = r_{t} + 2Ar\,.
\ee
Several equations (and their variants) fit into this scheme\,: see Table 1 for a list of $q,\,r,\,A,\,B$ and $C$. 
In a separate development, it was shown independently by Flaschka (1974) and Manakov (1975) that the Lax pair formulation can be extended to discrete systems when they came up with a Lax pair for what is known as the Toda lattice (Toda 1967). Fordy and Gibbons (1980) also showed that there exists an integrable Toda-like extension of the Klein-Gordon equation. Over the decades these developments provoked the writing of literally thousands of papers on the properties of solutions of this class of PDEs, including a series of textbooks\,: see, for example, Ablowitz and Segur (1981), Dodd \textit{et al} (1982), Novikov, Manakov, Pitaevskii and Zakharov (1984) and Newell (1985).
\par\smallskip\noindent
\begin{table}
{\tiny
\begin{tabular}{||c|c|c|c|c|c|c||}
	\hline
Name & PDE & $q$ & $r$ & $A$ & $B$ & $C$\\\hline\hline
NLS  & $iq_{t} + q_{xx} \mp 2 q|q|^{2} = 0$ & $q$ & $\pm q^{*}$ & $-2i\lambda^{2} \mp i |q|^{2}$ & $2\lambda q + iq_{x}$ 
& $\pm(2\lambda q^{*} - iq_{x}^{*})$\\\hline\
mKdV & $q_{t} \mp 6q^{2}q_{x} + q_{xxx}=0$ & $q$  & $\pm q$ & $-4i\lambda^{3} - 2i\lambda qr + rq_{x}-qr_{x}$ & 
$-q_{xx} + 2i\lambda q_{x} + 2q^{2}r + 4q\lambda^{2}$ & $-r_{xx} - 2i\lambda r_{x} + 2q r^{2} + 4r\lambda^{2}$\\\hline
SIT & $E_{xt} = EN$\,;~$2N_{x} = -(|E|^{2})_{t}$ & $E/2$ & $-E^{*}/2$ & $-N/4i\lambda$ & $E_{t}/4i\lambda$ & 
$E_{t}^{*}/4i\lambda$\\\hline
SGE & $\phi_{xt} = \sin\phi$ & $\phi_{x}/2$ & $-\phi_{x}/2$ & $-\cos\phi/4i\lambda$ & $\phi_{xt}/4i\lambda$ & $\phi_{xt}/4i\lambda$\\\hline
2nd HR & $P_{x} = Q P^{*}$\,;~$Q_{t} = P^{2}$ & $Q$ & $Q^{*}$ & $- |P|^{2}/2i\lambda$ & $P^{2}/2i\lambda$ & $-P^{*2}/2i\lambda$\\
\hline
\end{tabular}}
\caption{\scriptsize For the SIT and sine-Gordon equations (SGE) see Gibbon, James and Moroz (1979) and references therein\,; for the equations of 2nd harmonic resonance (2nd HR) see Kaup (1978).}
\end{table}


\section{\large The Painlev\'e property}

Martin had an enduring interest in the six Painlev\'e second order ordinary differential equations, designated as PI - PVI 
(see Ince 1956)\footnote{This section was written by Nalini Joshi.}. These are stated in Table 2.  His interest began when 
they appeared as symmetry reductions of soliton equations, but when his relentless questions about their properties were unanswered he dissected and broke down each question into natural components and developed new methods to find answers. 
\par\vspace{-3mm}\noindent
\begin{table}
\bc
{\scriptsize
\begin{tabular}{||c||c||}\hline
PI & $w'' = 6w^{2}+t$\\\hline
PII & $w'' = 2w^{3} + tw + \alpha$\\\hline
PIII & $w'' = w^{'2}/w - w/t + (\alpha w^{2} + \beta)/t +\gamma w^{3} + \delta/w$\\\hline
PIV & $w'' = w^{'2}/2w + 3w^{3}/2 + 4tw^{2} + 2w(t^{2}-\alpha) + \beta/w$\\\hline
PV  &  $w'' = \left(\frac{1}{2w} + \frac{1}{w-1}\right)w^{'2} - \frac{w'}{t} 
+ \frac{(w-1)^{2}}{t^{2}w}\left(\alpha w^{2} + \beta\right) + \frac{\gamma w}{t} + \frac{\delta w(w+1)}{w-1}$\\\hline
PVI & $w'' = \shalf\left(\frac{1}{w} + \frac{1}{w-1} + \frac{1}{w-t}\right)w^{'2} - \left(\frac{1}{t} + \frac{1}{t-1} + \frac{1}{w-t}\right)w' 
+ \frac{w(w-1)(w-t)}{t^{2}(t-1)^{2}}\left(\alpha + \frac{\beta t}{w^{2}} + \frac{\gamma (t-1}{(w-1)^{2}}  + \frac{\delta t(t-1)}{(w-t)^{2}}\right)$\\
\hline
\end{tabular}}
\caption{\scriptsize  The six Painlev\'e transcendents in terms of the independent variable $t$. The primes represent derivatives with respect to $t$ while $\alpha$, $\beta$, $\gamma$ and $\delta$ are constants.}
\ec
\end{table}
\par\vspace{-1mm}\noindent
Consistent with much of Martin's other work, each approach he developed had a major influence on other progress in the field. His strategy was loosely grouped in three directions\,: (i) describing singular behaviours\,; (ii) finding analytic properties of solutions\,; and (iii) asymptotic analysis. The process was always the same\,: focus on an interesting question, try an 
idea, resolve any paradox, and then follow the logic until a destination appears. At each step, examples and simpler models shaped the search for answers. In the standard analysis of linear ordinary differential equations (ODEs), if we solve for the highest derivative, the problematic places for defining solutions become clear\,: these are the places where the coefficient functions are singular. Such singularities are called \textit{fixed} singularities in the sense that their locations are determined for all time by the equation. However, these are not the only possibilities for nonlinear ODEs. Consider, for example, the Riccati equation $w' = t - w^{2}$, which is linearized by 
\bel{lin_ricc}
w = y'(t)/y(t)\,\quad\Rightarrow\quad y'' = ty\,.
\ee
The linear ODE governing $y(t)$ is the classical Airy equation, with a general solution $y(t) = a Ai(t) + b Bi(t)$, where 
$a$ and $b$ are arbitrary constants. The solution $w(t)$ of the Riccati equation becomes singular where $y(t)$ vanishes, but 
the locations of these zeros are determined by initial conditions. In turn, these determine the constants $a$ and $b$, which 
are not visible in the ODE. These are movable singularities\,; i.e., ones whose locations change or ``move" with initial conditions. In the above example, the solution $w(t)$ has an infinite number of movable poles but, in general, the solution 
may be multi-valued around movable singularities.
\par\smallskip
This property was well known to mathematicians in the 19th century who, by 1905, had classified second-order nonlinear ODEs 
(under some mild conditions) with the property that all solutions should be single-valued around all movable singularities 
(Ince 1956). Of the resulting class of 50 second-order ODEs, only the six Painlev\'e equations were found to define new higher transcendental functions as solutions. For these equations, all movable singularities of all solutions turn out to be poles. This is now called the \textit{Painlev\'e property}.
\par\smallskip
It created enormous excitement when it was discovered in the late 1970s that all known ODE reductions of soliton equations 
have this property\footnote{The easiest connection is to consider the mKdV equation \eqref{kdv3} in similarity variable form 
$v(x,t) = t^{-1/3}w(\tau)$ with $\tau= x t^{-1/3}$ and then integrate. This turns into a scaled form of PII with $\tau$ 
standing for $t$ in the table.}. Ablowitz, Ramani and Segur (1978, 1980) conjectured that this would always be the case and thereby provided a widely used test for integrability. Weiss, Tabor and Carnevale (1983) showed how to apply the test directly to PDEs. Martin provided a simplification of the procedure, called a \textit{reduced ansatz}, from the very beginning (listed 
as a private communication in the most cited paper of the time), creating clarity around these concepts in ways that are 
typical of his generous fundamental contributions to the field. He did not seek authorship in these cases, but others included him anyway\,: during a visit to the USA in 1982, Jimbo and Miwa visited Martin in Princeton and asked him questions about this test, resulting in its first application to the self-dual Yang-Mills system (Jimbo, Kruskal and Miwa 1982)\,: see Mason and Woodhouse (1996) for further results on this topic. 
\par\smallskip
Thousands of papers on the Painlev\'e property followed these developments. Whenever Martin was approached, he provided ideas 
and resolutions to the most common paradoxes and contradictions in the applications of the test. An application to the one-dimensional anisotropic Heisenberg spin chain in a transverse magnetic field is noteworthy because Martin pointed out that singularities occur not only where solutions become unbounded but also where terms multiplying the highest derivative may vanish, leading to a loss of information in the equation (Daniel, Kruskal, Lakshmanan and Nakamura 1992).  
\par\smallskip
These examples led to Martin's broad understanding of singularity analysis, perhaps the deepest of anyone in the field, 
but he still felt that fundamental questions remained unanswered. The test for the Painlev\'e property only gives necessary conditions. How do we find a sufficient proof that an example of interest does have the Painlev\'e property? There were integrable examples in which solutions were multivalued around movable singularities. How do we test for those? If singularities provide a good enough characterization of integrability, how do we find their other properties, such as Hirota's bilinear forms 
(Hirota 1971)? How can we test higher order ODEs, such as Chazy's third-order ODE which has a movable natural barrier, rather than a localized singularity (Clarkson and Olver 1996)? 
\par\smallskip
Martin always had an affectionate view of the Harry Dym equation, as an integrable equation related to the KdV equation, but 
its solutions have branched movable singularities and, therefore, it fails the test for the Painlev\'e property at the first step. Martin also had a model in his mind of a distinction between integrability and non-integrability based on whether solutions existed with a dense set of values at any point in the phase plane. Combining these two ideas, Martin was led to a major extension of the Painlev\'e property, in joint work with Clarkson, that he called the {\textit{poly-Painlev\'e} test 
for integrability (Kruskal and Clarkson 1992). As usual, his key ideas relied on asymptotic analysis, in this case covering 
many singularities concurrently.  Martin's insights in these directions also led to singularity analysis of nonlinear ODEs 
with extended movable singularities, such as natural barriers (Joshi and Kruskal 1993).
\par\smallskip
Martin's questions were often based on a radical premise\,: it is never acceptable to have an adulatory reference to a celebrated proof without knowing how to prove it yourself. He worked on providing a sufficient proof of the Painlev\'e 
property that was simpler than the ones provided by classical mathematicians so that it could be extended to all modern integrable equations. With Joshi, he provided a direct proof of the Painlev\'e property of the Painlev\'e equations based 
on first principles (Joshi and Kruskal 1994). Martin constantly worked on improving and simplifying this proof until he 
passed away in 2006. 
\par\smallskip
At an August meeting in Potsdam (NY) in 1979, Jim Corones, a materials scientist from Iowa, suggested one evening in jest 
that the test for integrability required only a postcard\,: write the equation of interest on a postcard, send it to Ryogo Hirota in Japan and if, in time, he sent back a very long formula then the equation must be integrable! Hirota's insights 
relied on finding a bilinear form of the equation, and has been used extensively to find specific soliton solutions more 
easily (Hirota 1971). Martin always wanted to know how to find the bilinear form directly, without relying on Hirota's intuition, and realized that he could do so by converting all movable singularities to movable regular zeroes of solutions. 
He used this idea in joint work with Hietarinta (Hietarinta and Kruskal 1992) to find Hirota forms of the Painlev\'e equations. 
\par\smallskip
In the summer of 1982, Martin was inspired by the idea that the highly transcendental solutions of Painlev\'e equations 
should be described in a similar way to the traditional classical special functions. In particular, he heard a talk by 
Bryce McLeod who suggested that the connection problem for Painlev\'e transcendents should be tackled in a similar way 
to that for Airy functions, by following a large semi-circular path in the complex plane (Hastings and McLeod 1980). 
Carrying this out turned out to be no mean feat, because standard averaging and multiple-scales methods had to be extended 
in counter-intuitive ways. These methodological extensions were achieved in Joshi's PhD thesis, supervised by Martin, and connection results in the complex plane were obtained for the first and second Painlev\'e equations (Joshi and Kruskal 1988,
\,1992a,\,1992b). 
\par\smallskip
Near infinity, the solutions of these Painlev\'e equations are asymptotic to (scaled) elliptic functions, which reduce to 
power series expansions for certain initial conditions in some sectors of the complex plane. These power series are 
divergent and hide a small free parameter \textit{beyond all orders} of the expansion. This was familiar territory for 
Martin, who had already encountered asymptotics-beyond-all-orders in the calculation of adiabatic invariants in 
plasma confinement. At the Santa Barbara conference celebrating Martin's 60$^{\rm th}$ birthday (Campbell and 
Kruskal 1986) and in the program that followed, he worked with Segur to resolve a similar problem that arose in the 
study of crystal growth in two dimensions (Kruskal and Segur 1991) and in the study of breathers in a field model 
approximating the sine-Gordon equation (Segur and Kruskal 1987). These papers have had a lasting influence on 
the field. Later, he applied ideas from surreal number theory with Costin to revisit this problem for classes of 
nonlinear ODEs (Costin and Kruskal 1999).
\par\smallskip
In the 1990s, discrete integrable versions of the Painlev\'e equations were proposed. Although Martin published 
only one paper on the subject (Kruskal, Tamizhmani, Grammaticos and Ramani 2000), his influence of this fledgling 
field was evident. At a conference in Esterel, Quebec in 1994, at a time when the popular but unsettled test for the 
singularity confinement property was being proposed as a discrete Painlev\'e property, Martin pointed out that it 
was actually a test for \textit{well-posedness} in the discrete equations. This observation alone sharpened the thinking 
at the time and allowed the field to develop and grow to the exciting, mature field it is today.

\section{\large Some final remarks}

It is in the nature of science that most of its participants tend to specialize to such a degree that they forever sit 
in their own valleys and pan for specks of gold in the local river. Many never even raise their heads to observe 
the hills that flank their particular valley. Martin was one of the few who climbed those hills, explored other 
valleys, and realized that the scientific disciplines form an inter-locking jigsaw whose picture tells a much bigger 
story. Not only did he make major contributions to the areas sketched in this memoir, he also had a lifelong 
interest in what he called `asymptotology', which he defined as ``as the art of dealing with applied mathematical 
systems in limiting cases'' (Kruskal 1963). He referred to asymptotology as ``an art, at best a quasi-science, but 
not a science''. 
\par\smallskip
Those who attended any of the early soliton conferences will recall a strongly mixed set of participants\,: not only 
regular PDE applied mathematicians but plasma and  optical physicists, fluid dynamicists, gauge theorists, geometric 
analysts, meteorologists, water wave theoreticians and experimentalists, and algebraic geometers. It was a heady 
rainbow mix of scientific cultures whose members participated because they felt something new was happening. It 
is ironic that in a day when many funding agencies across the world now require evidence of `interdisciplinarity' in 
their proposals, a grant-less Martin was an early founder of this style of interdisciplinary science.  In his eyes,  
interdisciplinarity was neither an institutionalized posture nor a box-ticking exercise, but simply the way he worked. 
The forward-thinking amiability of many of those early meetings could largely be attributed to Martin's friendly, robust 
and generous personality. He was always ready to give more credit to others and take less 
himself. He would also pepper speakers with endless questions and ideas although, at times, it could be an un-nerving 
experience to have one's mind turned inside out under the glare of such a penetrating intellect. 
\par\smallskip
\S\ref{sect:plasma} and \S\ref{sect:rel}} make it clear that Martin's early work is stamped indelibly all over modern 
plasma physics and relativity. Fifty years after the Gardner, Greene, Kruskal and Miura (1967) paper, it is pertinent to 
ask how successful the search for integrable systems has been? As ever, one can take both a narrow and a wide view. The 
narrow view is that there appears to be only a handful of integrable systems of physical significance, and that these are 
mainly restricted to one spatial 
dimension although, within this small subset, it ought to be acknowledged that the properties of the NLS equation in fibre 
optics have had profound consequences in that science (Agrawal 2011). In the early days it was hoped that the local particle 
properties of soliton solutions might be found in fully 3D systems but that appears, so far, to have been a vain hope. The 
2D systems that are integrable, such as the Davey-Stewartson and Kadomtsev-Petviashvili equations (Benney and Roskes
1969, Davey and Stewartson 1974, Kadomtsev-Petviashvili 1970), yield solutions that are more like wave-fronts rather than 
localized humps of finite energy. Associated with many members of the finite set of integrable systems there are infinite 
hierarchies of PDEs with no apparent physical significance, although the history of science shows that we should keep an 
open mind. The wider and more generous view is that Martin was a leader in teaching us that the physical world should 
not be seen through linear eyes, with a few special nonlinear solutions tacked on, but should be recognized not only as 
inherently nonlinear but should also be explored with confidence. Indeed, many rich mathematical structures associated 
with integrability have been discovered that were undreamed of before. Much of the early work on algebraic geometry 
associated with the KdV equation on periodic boundary conditions (Dubrovin and Novikov 1974, Lax 1975, Krichever 
1977, Segal and Wilson 1985) flowed out and merged with the `geometry and physics' revolution that was occurring in 
parallel. 
\par\smallskip
In modern academic and corporate circles, much is made of that ephemeral quality called `leadership'. Unfortunately, it is 
increasingly viewed in narrow terms, such as individual's ability to raise money or lead a large group. Kruskal did neither 
yet his very obvious leadership qualities lay in the realm of ideas and shone through to all who knew him. He belonged to 
that fading generation of scientists, educated just after WW2, who founded the international science research system we 
know today. The continuing financial support required by this system obviously needs those who excel in managing the 
processes that are necessary in this sphere, but it also requires individuals who not only have the ability and the vision to 
conjure major new ideas but also have the inspirational qualities to disseminate them across far-flung boundaries. Martin 
Kruskal excelled at this. Seen in this light, his far-seeing contribution to the understanding of modern physics and applied 
mathematics has been immense.

\bibliographystyle{unsrt}


\bc
\textbf{\large Bibliography -- papers by Kruskal referred to in the text}
\ec
\scriptsize
\bit\itemsep -1mm

\item[1952] U. S. Atomic Energy Commission Report No. NYO-998 (PM-S-5).

\item[1954] U. S. Atomic Energy Commission Report No. NYO-6045 (PM-S-12).

\item[1954] (with Schwarzschild, M.) Some Instabilities of a Completely Ionized Plasma. 
\textit{Proc. Royal. Soc. Lond. A} \textbf{223}, No 1154, 348--360. 

\item[1960] Maximal extension of Schwarzschild metric. \textit{Phys. Rev.} \textbf{119}, 1743--1745.

\item[1963] Asymptotology, 
\textit{Proceedings of Conference on Mathematical Models on Physical Sciences. Englewood Cliffs}. Prentice--Hall, 17--48.

\item[1965] (with Zabusky N. J.) Interaction of `solitons' in a collisionless plasma and the recurrence
of initial states. \textit{Phys. Rev. Letts.} \textbf{15}, 240--243. 

\item[1967] (with Gardner, C. S., Greene, J. M. and Miura R. M.) Method for solving the Korteweg de Vries equation. 
\textit{Phys. Rev. Letts.} \textbf{19}, 1095--1097. 

\item[1968] (with Gardner, C. S. Greene, J. M. and Miura, R. M.)  Korteweg de Vries equation and generalizations, II\,: 
Existence of conservation laws and constants of motion. \textit{J. Math. Phys.} \textbf{9}, 1204--1209. 

\item[1970] (with Miura, R. M., Gardner, C. S. and Zabusky, N. J.) Korteweg de Vries equation and generalizations 
V\,: Uniqueness and non-existence of polynomial conservation laws. \textit{J. Math. Phys.} \textbf{11}, 952--960. 

\item[1974] (with Gardner, C. S., Greene, J. M. and Miura, R. M.) Korteweg de Vries equation and generalizations 
VI\,: methods of exact solutions. \textit{Comm. Pure App. Math.} \textbf{27}, 97--133. 

\item[1982] (with Jimbo, M. and Miwa, T.) Painlev\'e test for the self-dual Yang-Mills equation. \textit{Phys. Letts.} 
\textbf{A92}, 59--60.

\item[1986] (with Campbell, D. K.) \textit{Solitons and Coherent Structures}, in Proceedings of the Conference on Solitons 
and Coherent Structures Held at Santa Barbara, CA, 93106, USA, January 11-16, 1985. Number 1-3. Northern California.

\item[1987] (with Segur, H.) Nonexistence of small-amplitude breather solutions in $\phi^4$ theory.  \textit{Phys. Rev. 
Letts.}, \textbf{58}, 747--750.

\item[1988] (with Joshi, N.) An asymptotic approach to the connection problem for the first and the second 
Painlev\'e equations. \textit{Phys. Lett. A} \textbf{130} 129--137.

\item[1991] (with Segur, H.) Asymptotics beyond all orders in a model of crystal growth. \textit{Stud. App. Math.} \textbf{85}, 129--181. 

\item[1992] a) (with Clarkson, P. A.) The Painlev\'e-Kowalevski and Poly‐Painlev\'e Tests for Integrability. 
\textit{Stud. App. Math.} \textbf{86},  87--165.

b) (with Daniel M. and Lakshmanan, M. and Nakamura, K.) Singularity structure analysis of the continuum 
Heisenberg spin chain with anisotropy and transverse field\,: nonintegrability and chaos. \textit{J. Math. Phys.} 
\textbf{33}, 771--776.

c) (with Hietarinta, J.) Hirota forms for the six Painlev\'e equations from singularity analysis.  
pgs 175-185 in \textit{Painlev\'e Transcendents}, Springer USA.

d) (with Joshi, N.)  The Painlev\'e connection problem\,: an asymptotic approach. I. \textit{Stud. 
App. Math.} \textbf{86}, 315--376.

e) (with Joshi, N.) Connection results for the first Painlev\'e equation, pgs. 61-79 in \textit{Painlev\'e Transcendents}, Springer USA.

\item[1993] (with Joshi, N.) A local asymptotic method of seeing the natural barrier of the solutions of the Chazy 
equation. In \textit{Applications of analytic and geometric methods to nonlinear differential equations}, pages 331--339. Springer.

\item[1994] (with Joshi, N.)  A direct proof that solutions of the six Painlev\'e equations have no movable 
singularities except poles. \textit{Stud. App. Math.} \textbf{83}, 187--207.

\item[1997] (with Joshi, N. and Halburd, R.) Analytic and asymptotic methods for nonlinear singularity analysis\,: 
a review and extensions of tests for the Painlev\'e property, pps 171--205 in \textit{Integrability of nonlinear 
systems}, Springer, Berlin and Heidelberg.

\item[1999] (with Costin, O.) On optimal truncation of divergent series solutions of nonlinear differential systems. \textit{Proc. Royal Soc. Lond. A} \textbf{A455}, 1931--1956.

\item[2000] (with Tamizhmani, K. M., Grammaticos B. and Ramani, A.) Asymmetric discrete Painlev\'e equations. \textit{Regular 
and Chaotic Dynamics} \textbf{5}, 273--280.

\eit

\newpage
\bc
\textbf{\large Full Kruskal Bibliography}
\ec
\scriptsize
\bit\itemsep -1mm

\item[1952] U. S. Atomic Energy Commission Report No. NYO-998 (PM-S-5).

\item[1954] a) U. S. Atomic Energy Commission Report No. NYO-6045 (PM-S-12).

													 b) (with Schwarzschild, M.) Some Instabilities of a Completely Ionized Plasma. 
													\textit{Proc. Royal. Soc. Lond. A} \textbf{223}, No 1154, 348--360. 

\item[1957] (with Bernstein, I. B. and Greene, J. M.) Exact Nonlinear Plasma Oscillations. \textit{Phys. Rev.} 
\textbf{108}, 546--550.

\item[1958]	a) (with Johnson, J. L., Gottlieb, M. B. and Goldman, L. M.) Hydromagnetic instability in a stellarator. 
\textit{Phys. Fluids} \textbf{1}, 421--429.

											b)	(with Kulsrud, R. M.) Equilibrium of a magnetically confined plasma in a toroid. \textit{Phys. Fluids} \textbf{1}, 265--274.

											c) (with Oberman, C. R.) On the stability of plasma in static equilibrium. \textit{Phys. Fluids} \textbf{1}, 275--280.
											
											d) (with Bernstein, I. B., Frieman, E. A. and Kulsrud, R. M.) An Energy Principle for Hydromagnetic Stability Problems. \textit{Proc. Royal. Soc. Lond. A} \textbf{244}, No 1236, 17--40.

\item[1960] Maximal extension of Schwarzschild metric. \textit{Phys. Rev.} \textbf{119}, 1743--1745.

\item[1962]	
											a) Asymptotic theory of Hamiltonian and other systems with all solutions nearly periodic. 
											\textit{J. Math. Phys.} \textbf{3}, 806--828.
											b) (with Greene, J. M.,  Johnson, J. L. and Wilets, L.) Equilibrium and stability of helical 
											hydromagnetic systems. \textit{Phys. Fluids} \textbf{5}, 1063--1069. 	

\item[1963] Asymptotology, 
\textit{Proceedings of Conference on Mathematical Models on Physical Sciences} Englewood Cliffs, NJ, Prentice--Hall, 17--48.

\item[1964] a)	(with Gardner, C. S.) Stability of plane magnetohydrodynamic shocks. \textit{Phys. Fluids} \textbf{7}, 700--706.

											b) (with Zabusky, N. J.) Stroboscopic-perturbation procedure for treating a class of nonlinear wave equations. 
														\textit{J. Math. Phys.} \textbf{5}, 231--244. 	

											c) (with Bernstein, I. B.) Runaway electrons in an ideal Lorentz plasma \textit{Phys. Fluids} \textbf{7} 407--418. 	

\item[1965] a)	(with Cohen, I. M.) Asymptotic theory of the positive column of a gas discharge. \textit{Phys. Fluids} 
\textbf{8}, 920--934. 

											b) (with Zabusky, N. J.) Interaction of `solitons' in a collisionless plasma and the recurrence of initial states. \textit{Phys. Rev. Letts.} \textbf{15}, 240--243. 

\item[1966]	a) (with Zabusky, N. J.) Exact invariants for a class of nonlinear wave equations. \textit{J. Math. Phys.} 
\textbf{7}, 1256--1267. 	

											b)	(with Northrop, T. G. and Liu, C. S.) First correction to the second adiabatic invariant of charged-particle motion. \textit{ Phys. Fluids} \textbf{9}, 1504--1513. 

											c)	(with Orszag, S. A.) Theory of turbulence. \textit{Phys. Rev. Letts.} \textbf{16} 441--444.

\item[1967] (with Gardner, C. S., Greene, J. M. and Miura, R. M.) Method for solving the Korteweg de Vries equation. 
\textit{Phys. Rev. Letts.} \textbf{19}, 1095--1097. 

\item[1968] a) (with Orszag, S. A.) Formulation of the theory of turbulence. \textit{Phys. Fluids} \textbf{11}, 43--60. 	

											b) (with Gardner, C. S., Greene, J. M. and Miura, R. M.)  Korteweg de Vries equation and generalizations, II\,: 
											Existence of conservation laws and constants of motion. \textit{J. Math. Phys.} \textbf{9}, 1204--1209. 

\item[1969]	(with Valeo, E. J. and Oberman, C.) Property of large-amplitude steady electrostatic waves. \textit{Phys. Fluids} 
\textbf{12}, 1246, 1969. 

\item[1970] a)	(with G. V. Ramanathan and J. M. Dawson) Entropy principle for the derivation of a new cluster expansion. 
\textit{J. Math. Phys.} \textbf{11} 339--342. 

b) (with Miura, R. M., Gardner, C. S. and Zabusky, N. J.) Korteweg de Vries equation and generalizations V\,: Uniqueness and 
non-existence of polynomial conservation laws. \textit{J. Math. Phys.} \textbf{11}, 952--960. 

\item[1971] (with Miller, J. G.	 and Godfrey, B. B.) Taub-nUT (Newman, Unti, Tamburino) metric and incompatible extensions. 
\textit{Phys. Rev. D} \textbf{4}, 2945--2948.

\item[1972]	(with Bateman, G.) Linear time-dependent Vlasov equation\,; Case-Van Kampen modes. \textit{Phys. Fluids} 
\textbf{15}, 277--283. 	

\item[1973]	(with Miller, J. G.) Extension of a compact Lorentz manifold. \textit{J. Math. Phys.} \textbf{14}, 484--485. 	

\item[1974] (with Gardner, C. S., Greene, J. M. and Miura, R. M.) Korteweg de Vries equation and generalizations 
VI\,: methods of exact solutions. \textit{Comm. Pure App. Math.} \textbf{27}, 97--133. 

\item[1979] a)	(with Ablowitz, M. J. and Ladik, J. F.) Solitary wave collisions. \textit{SIAM J. Appl. Math.} \textbf{36}, 
428--437. 

											b) (with Fisch, N. J.) Separating variables in two-way diffusion equations. \textit{J. Math. Phys.} 
											\textbf{21}, 740--750. 	

											c)	(with Bock, T. L.) A two-parameter Miura transformation of the Benjamin-Ono equation. 
											\textit{Phys. Letts. A} \textbf{74}, 173--176. 

\item[1982] (with Jimbo, M. and Miwa, T.) Painlev\'e test for the self-dual Yang-Mills equation. 
\textit{Phys. Letts.} \textbf{A92}, 59--60.

\item[1984] (with	Peyrard, M.) Kink dynamics in the highly discrete sine-Gordon system. \textit{Phys. D\,: Nonlin. Phen.} 
\textbf{14}, 88--102. 

\item[1986] (with Campbell, D. K.) \textit{Solitons and Coherent Structures}, in Proceedings of the Conference on Solitons 
and Coherent Structures Held at Santa Barbara, CA, 93106, USA, January 11-16, 1985. Number 1-3. Northern California.

\item[1986]	(with Joshi, N.) The connection problem for Painlev\`e transcedents. \textit{Physica D\,: Nonlin. Phen.} \textbf{18}, 215--216. 

\item[1987] (with Segur, H.) Non-existence of small-amplitude breather solutions in $\phi^4$-theory.  \textit{Phys. Rev. Letts.}, 
\textbf{58}, 747--750.

\item[1988] (with Joshi, N.) An asymptotic approach to the connection problem for the first and the second Painlev\'e equations. 
\textit{Phys. Lett. A} \textbf{130} 129--137.

\item[1989] (with Clarkson, P. A.) New similarity reductions of the Boussinesq equation. \textit{J. Math. Phys.} \textbf{30}, 2201-2213.

\item[1991] (with Segur, H.) Asymptotics beyond all orders in a model of crystal growth. \textit{Stud. App. Math.} \textbf{85}, 129--181. 

\item[1992] a) (with Clarkson, P. A.) The Painlev\'e-Kowalevski and Poly‐Painlev\'e Tests for Integrability. 
\textit{Stud. App. Math.} \textbf{86},  87--165.

b) (with Daniel, M. and  Lakshmanan, M. and Nakamura, N.) Singularity structure analysis of the continuum 
Heisenberg spin chain with anisotropy and transverse field\,: nonintegrability and chaos. \textit{J. Math. Phys.}  
\textbf{33}, 771--776.

c) (with Hietarinta, J.) Hirota forms for the six Painlev\'e equations from singularity analysis.  
pgs. 175--185 in \textit{Painlev\'e Transcendents}, Springer USA.

d) (with Joshi, N.)  The Painlev\'e connection problem\,: an asymptotic approach. I. \textit{Stud. 
App. Math.} \textbf{86}, 315--376.

e) (with Joshi, N.) Connection results for the first Painlev\'e equation, pgs. 61-79 in \textit{Painlev\'e Transcendents}, Springer USA.

\item[1993] (with Joshi, N.) A local asymptotic method of seeing the natural barrier of the solutions of the Chazy equation. 
In \textit{Applications of analytic and geometric methods to nonlinear differential equations}, pages 331--339. Springer.

\item[1994] (with Joshi, N.)  A direct proof that solutions of the six Painlev\'e equations have no movable singularities 
except poles. \textit{Stud. App. Math.} \textbf{83}, 187--207.

\item[1996]	(with Costin, O.) Optimal uniform estimates and rigorous asymptotics beyond all orders for a class of ordinary differential equations.  \textit{Proc. Royal Soc. A} \textbf{452}, 1057--1085. 

\item[1997] (with Joshi, N. and Halburd, R.) Analytic and asymptotic methods for nonlinear singularity analysis\,: 
a review and extensions of tests for the Painlev\'e property, pps 171--205 in \textit{Integrability of nonlinear systems}, Springer, 
Berlin and Heidelberg.

\item[1999] (with Costin, O.) On optimal truncation of divergent series solutions of nonlinear differential systems. \textit{Proc. Royal Soc. Lond. A} \textbf{A455}, 1931--1956.

\item[2000] (with Tamizhmani, K. M., Grammaticos, B. and Ramani, A.) Asymmetric discrete Painlev\'e equations. 
\textit{Regular and Chaotic Dynamics} \textbf{5}, 273--280.

\item[2003]	(with Costin, R. D.) Nonintegrability criteria for a class of differential equations with two regular singular points. \textit{Nonlinearity} \textbf{16}, 1295--1317.

\item[2004] (with Costin, O. and Dupaigne, L.) Borel summation of adiabatic invariants. \textit{Nonlinearity} \textbf{17},1509--1519.  

\item[2005]	(with Costin, O.) Analytic methods for obstruction to integrability in discrete dynamical systems. \textit{Comm. 
Pure Appl. Math.} \textbf{58}, 723--749. 

\eit
\end{document}